  \documentstyle[twocolumn,prl,aps,psfig,epsf,amssymb,amsfonts]{revtex}
\begin{document}
%
%
\renewcommand{\Re}{\operatorname{Re}}
\renewcommand{\Im}{\operatorname{Im}}
\newcommand{\Tr}{\operatorname{Tr}}
\newcommand{\sign}{\operatorname{sign}}
\newcommand{\dd}{\text{d}}
\newcommand{\q}{\boldsymbol q}
\newcommand{\p}{\boldsymbol p}
\newcommand{\rr}{\boldsymbol r}
\newcommand{\pp}{p_v}
\newcommand{\vv}{\boldsymbol v}
\newcommand{\I}{{\rm i}}
\newcommand{\pphi}{\boldsymbol \phi}
\newcommand{\ds}{\displaystyle}
\newcommand{\be}{\begin{equation}}
\newcommand{\ee}{\end{equation}}
\newcommand{\bea}{\begin{eqnarray}}
\newcommand{\eea}{\end{eqnarray}}
\newcommand{\Acl}{{\cal A}}
\newcommand{\Rcl}{{\cal R}}
\newcommand{\Tcl}{{\cal T}}
\newcommand{\Tmin}{{T_{\rm min}}}
\newcommand{\Toff}{{\langle \delta T \rangle_{\rm off} }}
\newcommand{\Roff}{{\langle \delta R \rangle_{\rm off} }}
\newcommand{\RoffI}{{\langle \delta R_I \rangle_{\rm off} }}
\newcommand{\RoffII}{{\langle \delta R_{II} \rangle_{\rm off} }}
\newcommand{\dg}{{\langle \delta g \rangle_{\rm off} }}
\newcommand{\rd}{{\rm d}}
\newcommand{\br}{{\bf r}}
\newcommand{\la}{\langle}
\newcommand{\ra}{\rangle}

\twocolumn[\hsize\textwidth\columnwidth\hsize\csname @twocolumnfalse\endcsname
%
%
\draft


\title{Semiclassical Construction of Random Wave Functions
                   for Confined Systems}

\author{Juan Diego Urbina and Klaus Richter}

\address{
 Institut f\"ur Theoretische Physik, Universit\"at Regensburg, 
         93040 Regensburg, Germany}

\date{\today}
\maketitle

\begin{abstract}
We develop a statistical description of chaotic wavefunctions in closed systems
obeying arbitrary boundary conditions by combining a semiclassical expression for the
spatial two-point correlation function with a treatment of 
eigenfunctions as Gaussian random fields. Thereby we generalize Berry's isotropic random
wave model by incorporating confinement effects through classical paths reflected at the boundaries. Our approach allows to explicitly calculate highly non-trivial statistics, such
as intensity distributions, in terms of usually few short orbits,
depending on the energy window considered. 
We compare with numerical quantum results for the Africa billiard
and derive non-isotropic random wave models for other prominent confinement geometries. 
\end{abstract}


%

]


 \narrowtext

In mesoscopic quantum systems at low temperatures and far away from any phase 
transition,  many of the relevant physical phenomena can be described in the 
mean field approximation. In this scheme the excitations of the system are 
considered as a set of independent quasi particles with energies in 
a small range around the Fermi energy, which for many-particle systems is much larger than  
the single-particle ground state energies \cite{Im}. 
Hence, in this semiclassical regime, characterized by (Fermi) wave lengths 
considerably smaller than the system size, challenges to theory are posed owing to 
the arising complexity of the single-particle wave functions involved.
In view of the correspondence principle their structures depend sensitively 
on phase space properties of the corresponding classical system \cite{gut}. This
has called for an increasing theoretical investigation of 
statistical properties of eigenstates \cite{cambridge} since the seminal works
by Berry \cite{berr1} and by McDonald and Kaufman \cite{mac}. 
This is of more than theo\-re\-ti\-cal interest as fluctuations of wave
function amplitudes govern a variety of physical processes such as, e.g.,
photodissociation of molecules and the measured statistics of conductance peaks 
\cite{coulomb-blockade-exp} in the Coulomb blockade regime 
\cite{note-CB}. Moreover, advances in scanning probe tech\-niques and microwave experiments allow to directly uncover the spatial structure of
waves on mesoscopic scales\cite{sto}.

To mimic the statistical properties of wavefunctions in classically chaotic
quantum systems, Berry conjectured \cite{berr1} that chaotic wavefunctions behave as Gaussian 
random fields, and arguments coming from semiclassics \cite{berr1}, 
quantum ergodicity \cite{uzy1}, and information theory \cite{har2} support this 
Gaussian hypothesis. When supplemented with a Bessel-type spatial two-point correlation 
function, the resulting theory is known as Berry's Random Wave Model (RWM), 
since it is equivalent to consider the wavefunction as a random superposition of plane 
waves with locally fixed wavenumber magnitude. The RWM provides universal, 
system-independent results consistent with random
matrix theory. It constitutes the most widely used statistical 
description of chaotic eigenfunctions, as it has been extremely successful in 
predicting bulk or spatially averaged quantities. However, obviously, the 
RWM does not account for effects of confinement potentials which 
pose additional constraints to the wave functions, 
reducing their randomness particularly in the spatial region close to the boundaries.
This fact strongly diminishes the range of applicability of the usual RWM, since
in many experimental situations the behavior of the wave function close to the boundary 
is particularly relevant (e.g.\ when measuring tunnel rates, the local density
of states at surfaces or boundaries, or the conductance by attaching leads).
Hence, very recently several papers appeared, where boundary effects have been 
incorporated into RWM approaches, however for very specific geometries
\cite{berr2,berr3,hell,hell2} only or in a qualitative way\cite{denn}.

In this Letter we construct a RWM which allows to incorporate
boundary effects of arbitrary confinements including Dirichlet-, Neumann-, and 
mixed boundary conditions in both billiard and smooth systems.
We combine the Gaussian conjecture for eigenfunction statistics 
with a semiclassically exact representation of the spatial two-point correlation function. 
This enables us to account for confinement-induced random wave
correlations in terms of usually few classical paths, generalizing and improving ideas presented 
in \cite{berr1,prig,sied1,sied2}. We illustrate the generality and strength of our 
technique for different systems including those treated in \cite{berr2,berr3,hell,hell2}.

{\it Defining the ensemble.} We focus on two-dimensional clean, 
closed systems with time reversal symmetry\cite{note1}.
We consider energy averages over a set of $N_{W}$ normalized solutions 
${\psi_{n}(\vec{r})}$ of the Schr\"odinger equation with non-degenerate 
eigenvalues ${E_{n}}$ lying inside an interval $W=[e-\delta e / 2,e+\delta e / 2]$. 
We assume $\delta e / e \ll 1$,
which can always be achieved in the semiclassical limit we are interested in.
Considering such energy averages is standard for disorder-free mesoscopic
systems as it allows for random matrix approaches \cite{bee}. Moreover, experiments often involve averages 
over finite energy windows\cite{exp2}.
In particular, the energy-averaged eigenfunction intensity to be considered is 
proportional to the local density of states, relevant to many experiments
such as photoabsorption, quantum transport, and ionization processes.

At a fixed position $\vec{r}=(x,y)$ we will probe wave function amplitudes
by means of a function  
$F(u_{n})=F(\psi_{n}(\vec{r}))$ which fluctuates when varying
$E_{n}$ and the corresponding state $\psi_{n}$ inside $W$. We define the spectral average of $F$ at $\vec{r}$ as ${\cal F}(\vec{r}) \equiv \frac{1}{N_{W}}\sum_{E_{n} \in W}F(\psi_{n}(\vec{r}))$. A typical example is the distribution of intensities,
$I(w;\vec{r}) \equiv \frac{1}{N_{W}}\sum_{E_{n} \in W}\delta(w\!-\!|\psi_{n}(\vec{r})|^2)$. 
The definition is easily generalized to higher-order statistics such as
the spatial correlation of the intensity distribution,
$Y(w_{1},w_{2};\vec{r}_{1},\vec{r}_{2}) \equiv
 \frac{1}{N_{W}}\sum_{E_{n} \in W}\delta(w_{1}-|\psi_n(\vec{r}_{1})|^2)\
\delta(w_{2}-|\psi_n(\vec{r}_{2})|^2)$, and to
functions $F(\vec{u})$ depending not only on the eigenfunctions 
but also on their derivates of any order:
\begin{equation}
\label{eq:FM}
{\cal  F}(\vec{r}_{1},\ldots, \vec{r}_{M})  
\equiv \frac{1}{N_{W}}\sum_{E_{n} \in W}F(u^{1}_{n}(\vec{r}_{1}),..,u^{M}_{n}(\vec{r}_{M}))
\end{equation}
where $u^{\alpha}_{n}(\vec{r}_{i})=\partial^{l_{\alpha}}_{x_{i}} 
\partial^{m_{\alpha}}_{y_{i}}\psi_{n}(x_{i},y_{i})$ 
with integers $l_{\alpha},m_{\alpha}$. If there are $J$ different positions 
among the set ${\vec{r}_{1},\ldots,\vec{r}_{M}}$, we call  ${\cal F}(\vec{r}_{1},\ldots, 
\vec{r}_{M}) $ a $J$-point statistics.  
In this paper, a central quantity is the two-point correlation function
\begin{equation}
\label{eq:R}
R(\vec{r}_{i},\vec{r}_{j}) \equiv 
\frac{1}{N_{W}}\sum_{E_{n} \in W}\psi_{n}(\vec{r}_{i})\psi_{n}(\vec{r}_{j}) \; ,
\end{equation}
since the average of any expression bilinear in the wavefunction can be 
expressed through this correlation.

{\it The Gaussian conjecture.} Introducing the joint probability distribution $ P(\vec{u})=\frac{1}{N_{W}}\sum_{E_{n} \in W}\delta(\vec{u}-\vec{u}_{n}) $ the statistics (\ref{eq:FM})
can be cast into the more familiar form 
${\cal F}(\vec{r}_{1}, \ldots , \vec{r}_{M}) =\int_{-\infty}^{\infty}F(\vec{u})P(\vec{u})d\vec{u}$. 
The {\it Gaussian conjecture for the statistics of eigenfunctions of classically chaotic 
quantum systems} claims that the energy ensemble is described as a Gaussian 
stationary process. More precisely, this means to assume (in the weak sense) $P(\vec{u})=(2 \pi)^{-M/2} \left({\rm det}{\bf C}\right)^{-1/2} {\rm exp}
\left(-\frac{1}{2}\vec{u}({\bf C}^{-1})\vec{u}\right)$, where the correlation matrix ${\rm \bf C}={\rm \bf C}(\vec{r}_{1},\ldots,\vec{r}_{J})$ 
has entries $c_{\alpha,\beta}=\frac{1}{N_{W}}\sum_{E_{n} \in W}u^{\alpha}_{n}u^{\beta}_{n}$. 
Since all these entries consist of averages over quantities bilinear 
in the eigenfunctions, the knowledge of the two-point correlation function (\ref{eq:R})
completely determines, under the Gaussian assumption, 
the matrix $\bf C$ and all statistical properties.

Applying this approach to the intensity distribution $I(w;\vec{r})$, the 
 matrix  $\bf C$ reduces to a single entry $c_{1,1}=R(\vec{r},\vec{r})$.
Using the above expression for $P(u)$ we find 
\begin{equation}
\label{eq:I}
I(w;\vec{r})=\frac{1}{\sqrt{w R(\vec{r},\vec{r})}}{\rm exp}
\left(-\frac{w}{2 R(\vec{r},\vec{r})}\right) \; .
\end{equation}
Due to the presence of the boundary,
$R(\vec{r},\vec{r})$ will  generally depend on $\vec{r}$ (as will be discussed in Fig.\ 1).
This constitutes a {\em non-isotropic} generalization of the (isotropic) Porter-Thomas distribution, 
given by $R(\vec{r},\vec{r})=$ const.

The correlation of the intensity distribution, $Y$, involves a $2\times 2$ correlation matrix 
with elements $c_{i,j}=R(\vec{r}_{i},\vec{r}_{j})$. The Gaussian integrals then give
\begin{eqnarray}
\label{eq:Y}
&& Y(w_{1},w_{2};\vec{r}_{1},\vec{r}_{2})=\frac{1}{2 \pi 
\sqrt{w_{1} w_{2} {\rm det}{\bf C}}} \times \nonumber \\
& \times & {\rm cosh}\left(\frac{\sqrt{w_{1}w_{2}}c_{1,2}}{{\rm det}{\bf C}}\right)
 {\rm exp}\left(-\frac{c_{1,1}w_{2}+c_{2,2}w_{1}}{2{\rm det}{\bf C}}\right) \; ,
\end{eqnarray}
which is the non-isotropic generalization of the distribution studied in \cite{prig,sied1}.

{\it Semiclassical construction of the correlation matrix.} 
The above scheme critically depends  
on how precisely $R(\vec{r}_i,\vec{r}_j)$ can be calculated. This is a serious issue 
in the theory of chaotic quantum systems where no analytical expressions for the 
eigenfunctions exist, and approximate methods are required. It turns out convenient 
to express $R(\vec{r}_i,\vec{r}_j)$ through the Green function 
$G(\vec{r}_i,\vec{r}_j;E+i 0^+)$, 
\begin{equation}
\label{eq:ImG}
R(\vec{r}_i,\vec{r}_j)= \frac{1}{\pi} 
\frac{1}{N_{W}}\int_{e-\delta e/2}^{e+\delta e/2}{\rm Im} \
G(\vec{r}_i,\vec{r}_j;E+i 0^{+}) dE ,
\end{equation}
since a variety of approximations exists for $G$.

We start from the the exact multiple reflection expansion
of the Green function \cite{bal} and consider the two leading terms,
$G\simeq G^{(0)}+G^{(1)}$, to calculate $R(\vec{r}_i,\vec{r}_j)$.

The term $G^{(0)}$ denotes the contribution from the direct path
joining $\vec{r}_{i}$ and $\vec{r}_{j}$.
The corresponding isotropic contribution $R^{\rm is}(\vec{r}_{i},\vec{r}_{j})$ to $R$
can be calculated directly from Eq.~(\ref{eq:ImG}) by means of the
short-time propagator for direct paths.
For small distances $q=|\vec{r}_{i}-\vec{r}_{j}|$ \cite{pot} it
is evaluated at the mean potential $V(\vec{Q})$ 
for a local wave number $\hbar k=[2m(e-V(\vec{Q}))]^{1/2} $ with mass $m$ and
 $\vec{Q}=(\vec{r}_{i}+\vec{r}_{j})/2$:
\begin{equation}
\label{eq:Ris}
R^{is}(\vec{r}_{i},\vec{r}_{j}) = 
\frac{m\delta e}{2\pi \hbar^{2} N_{W}}J_{0} 
\left(kq\right)\Gamma\left(\frac{kq\delta e}{e}\right) \; .
\end{equation}
Here, $\Gamma(x)=\sin x/x$ is a window function; 
$\Gamma(x)\!=\!1$ corresponds to Berry's celebrated result \cite{berr1,sied2,cay}
for the isotropic RWM.
By choosing $\delta e= \hbar/\tau_{l}$ with $\tau_{l}$ the ballistic 
time scale associated with the system size $l$, one obtains $\Gamma(2q/l)$ \cite{yo}, i.e.\
the correlation function is suppressed on distances of the order of the 
system size\cite{note-finite-size}. 

The second term, $G^{(1)}$, represents all quantum paths between
$\vec{r}_{i}$ and $\vec{r}_{j}$ hitting the boundary once (including non-specular
reflections). 

The power of the representation (\ref{eq:ImG}) for the correlator $R(\vec{r}_i,\vec{r}_j)$
combined with the Green function expansion
is demonstrated for the Africa billiard\cite{Africa} depicted in the left inset of Fig.~1. 
The numerical evaluation of $R(\vec{r},\vec{r})$ 
(see Fig.~1) and $R(0,\vec{r})$ (right inset in Fig.~1) 
within this approximation is extremely fast and the results (thin lines) show considerable
agreement with numerically exact, but time consuming quantum mechanical reference 
calculations (symbols). 
The boundary effects (e.g, the oscillations in $R(\vec{r},\vec{r})$)
are adequately incorporated in the one-bounce treatment, but evidently beyond the 
range of applicability of the isotropic RWM, Eq.~(\ref{eq:Ris}) (dashed lines),
which yields, e.g., $R(\vec{r},\vec{r}) =$ const.

In the semiclassical limit the terms in the multiple reflection expansion can be 
further approximated by the respective semiclassical Green function \cite{gut}
 $G^{\rm sc}(\vec{r}_{i},\vec{r}_{j};E) = (i \hbar \sqrt{2 \pi i \hbar})^{-1}
 \sum_{\gamma}|D_{\gamma}|^{1/2}
\exp\left(\frac{i}{\hbar}S_{\gamma}-i\mu_{\gamma} \frac{\pi}{2}\right)  $ where $\gamma$ now labels the {\it classical} paths joining $\vec{r}_{i}$ with $\vec{r}_{j}$.
$D_{\gamma}$ and $\mu _{\gamma}$ are smooth  
classical quantities, and $S_{\gamma}(\vec{r}_{i},\vec{r}_{j};E)=
\int_{\gamma} \vec{p}\cdot d\vec{q}$ is the classical action along the path. 
For energy windows satisfying $\delta e / e \ll 1$,
the energy integral (\ref{eq:ImG}) then yields
the two-point correlation function
\begin{eqnarray}
\label{eq:Rsc}
R(\vec{r}_{i},\vec{r}_{j}) \simeq \frac{2 m 
\delta e}{(2 \pi \hbar)^{3/2}N_{W}}\sum_{\gamma}\Gamma\left(\frac{T_{\gamma}}{\tau_{W}}\right)
|D_{\gamma}|^{1/2} \times
&& \\ \times \ {\rm cos}\left[S_{\gamma}(\vec{r}_{i},\vec{r}_{j};e)/\hbar -\mu_{\gamma}
\pi/2\right] \;  && \nonumber
\end{eqnarray}
in terms of classical paths.
In Eq.~(\ref{eq:Rsc}), $T_{\gamma}$ is the traversal time of path $\gamma$, and 
$\tau_{W}=2 \hbar/\delta e$ is a characteristic cut-off time 
associated with the energy window $W$. 
Eq.\ (\ref{eq:Rsc}), representing a generalization of the correlator $R$
conjectured in Ref.\ \cite{sied2} (given by taking $\Gamma(x)=1$), incorporates
three significant advantages:  
First, it is semiclassically exact. Second, it allows to appropriately
describe the statistics for a chosen energy window by controlling the longest
path to be included via $\tau_{W}$, while this time scale is missing in the case $\Gamma(x)=1$.
Third, most importantly, it is compatible  \cite{explain} with the definition of a correlation,
(\ref{eq:R}), contrary to the correlation used in \cite{sied2}.

\begin{figure} 
\begin{center}
\psfig{figure=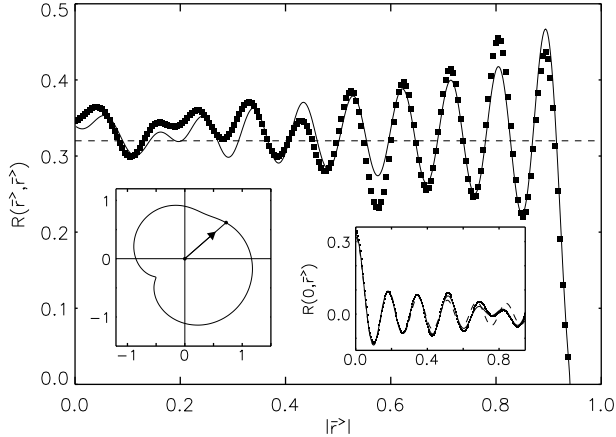,width=0.45\textwidth,height=0.32\textwidth,angle=0}
 \end{center}
 \vspace{-6mm}
\caption{Two-point correlation function $R(\vec{r},\vec{r})$ and $R(0,\vec{r})$ (right inset)
for $\vec{r}$ pointing along the line indicated in the Africa billiard (left inset). 
The symbols mark numerical quantum results for $R$,  Eq.~(\ref{eq:R}) 
 \protect\cite{note-qm}.
The thin lines depict the semi-quantum prediction employing Eq.~(\ref{eq:ImG}) 
where the Green function is approximated by a sum over 
paths, including diffraction effects, with at most one reflection at the boundary. 
The dashed lines show the isotropic RWM result (\ref{eq:Ris}).}
\end{figure}

The isotropic correlation $R^{\rm is}$, Eq.~(\ref{eq:Ris}), turns out to be extremely 
robust with respect to an additional spatial average.
To see this we note that for fixed  $\vec{Q}$ the integration over the relative position $\vec{q}$ in any small region will contain 
the continuous set of paths joining $\vec{r}_{i}$ with $\vec{r}_{i}+\vec{q}$ directly 
and the contribution from non-direct paths being isolated in chaotic systems. 
Hence in the semiclassical limit the spatial integration over the continuous set of direct 
paths yields the dominant contribution which coincides with the isotropic result.

On the contrary, for a pure energy average the contribution 
from  non-direct paths to the correlation is of the same semiclassical order 
than that from direct paths. However, the width $\delta e$ (corresponding to the number 
 $N_{W}$ used to define the ensemble) determines the maximum length of the
non-direct paths contributing to the correlation function. The major step beyond 
the isotropic case is then to consider an energy window such that only the direct 
and shortest non-direct paths significantly contribute to the correlation 
function, i.e.\ a situation which is also particularly experimentally relevant.

To this end we need to specify the non-direct paths more precisely.
In billiard systems the first non-direct con\-tribution to $R$ is given by a
sum $\sum_{p}R^{(p)}(\vec{r}_{i},\vec{r}_{j})$ over usually
few classical trajectories $p$ hitting the boundary once.  
For given initial and final positions $\vec{r}_{i}$, $\vec{r}_{j}$ each one-bounce 
path $p$ is uniquely characterized by the position $\vec{r}_{p}$ where it is reflected.
The path length is $L_{p}=L_{ip}+L_{jp}$ with 
$L_{ip}=|\vec{r}_{i}-\vec{r}_{p}|$, $L_{jp}=|\vec{r}_{j}-\vec{r}_{p}|$. 
Denoting by $\kappa_{p}$ and  $\theta_{p}$ the local boundary curvature and reflection
angle at $\vec{r}_{p}$, a simple calculation yields for each path
\begin{eqnarray}
\label{eq:Rp}
R^{(p)}(\vec{r}_{i},\vec{r}_{j}) & \simeq & 
\Gamma\left(\frac{kL_{p}\delta e}{e}\right)
\left |2\kappa_{p} \left( \frac{L_{ip}L_{jp}}{L_{p}{\rm cos} \theta_{p}}\right)-1\right
|^{-\frac{1}{2}} \times
 \nonumber \\  & \times &
\frac{1}{A\sqrt{2 \pi k L_{p}}}{\rm cos}\left(k L_{p}-\frac{\pi}{4}+\phi_{p}\right). 
\end{eqnarray}
to the correlation function.
Here $A$ is the billiard area, and $\phi_p$ takes into account the boundary conditions at the reflection point, 
as given e.g.\ in \cite{uzy2}.  

The function $R(\vec{r}_{i},\vec{r}_{j})=
R^{\rm is}(\vec{r}_{i},\vec{r}_{j})+\sum_{p}R^{(p)}(\vec{r}_{i},\vec{r}_{j})$,
together with the semiclassical expressions (\ref{eq:Ris}) and (\ref{eq:Rp}),
provide the entries for the correlation matrix $\bf C$, from which arbitrary
statistical measures (such as $I$, Eq.~(\ref{eq:I}), and $Y$, Eq.~(\ref{eq:Y})) 
for the wave functions can be deduced beyond the isotropic case. Moreover this semiclassical 
correlation yields closed analytical expressions for statistical quantities
for chaotic systems as the stadium-, cardiod-, or Sinai-billiard, 
since in these cases all the  parameters required 
are readily calculated from geometrical considerations. 

The preceding discussion is easily generalized to systems with Aharonov-Bohm
flux lines or with smooth boundary potentials, where the first non-direct 
contribution includes paths with one classical turning point.

{\it Non-isotropic random wave models: one path is enough.} 
In the following we demonstrate the power of the general semiclassical scheme outlined 
above by computing wave function correlators for selected, representative examples.
First we show the role of confinement effects by considering 
points $\vec{r}$ close to the boundary of a billiard system.
In \cite{berr2} this situation is treated by approximating the boundary by an 
infinite straight line $y=y_{0}$ and considering an ensemble of random superpositions of plane waves $\psi^{r}(\vec{r})$ satisfying the general, 
mixed boundary condition $(\partial_{y}\psi^{r}(\vec{r})\sin{a}+k\psi^{r}(\vec{r})\cos{a})|_{y=y_{0}}=0$. Here $a$ is a generally position-dependent parameter and $k$ is the local wavenumber. By ensemble average a variety of two-point correlations was derived in \cite{berr2} 
and used to calculate specific statistical
observables. To illustrate our method and for the sake of comparison we consider 
in detail just one such average, namely
$\langle \psi^{r}(\vec{r})\partial_{y}\psi^{r}(\vec{r})\rangle$. 
In terms of the two-point correlation function it reads
$\langle \psi(\vec{r})\partial_{y}\psi(\vec{r})\rangle=\frac{1}{2}(\partial_{y_{i}}\!+\!
\partial_{y_{j}})R(\vec{r}_{i},\vec{r}_{j})|_{y_{i}=y_{j}=y}$. 
Close to the boundary only the direct and the shortest non-direct path contribute. 
For mixed boundary conditions the ex\-tra phase $\phi_p$ in Eq.~(\ref{eq:Rp}) is 
given semiclassically\cite{uzy2} 
by $\phi_{p}=\pi-2\arctan\left(\tan a\cos\theta_{p}\right)$.
Substitution into  Eq.\ (\ref{eq:Rp})
gives the approximate correlation function close to an arbitrary boundary. 
To leading order in $k$ we get
\begin{eqnarray}
\label{eq:exam1}
\langle \psi(\vec{r})\partial_{y}\psi(\vec{r})\rangle & = & 
\Gamma\left(\frac{2kd(\vec{r})\delta e}{e}\right) \frac{1}{\sqrt{| 1-\kappa d(\vec{r})|}}
\times  \\
&\times & \frac{k}{A}\frac{1}{\sqrt{\pi k d(\vec{r})}}{\rm  sin}
\left(2kd(\vec{r})-2a-\frac{\pi}{4}\right) \; , \nonumber 
\end{eqnarray} 
where $d(\vec{r})$ is the shortest distance from $\vec{r}$ to the boundary. 
In the limit of flat boundaries or very short distances $d(\vec{r})$ the 
semiclassical results represent
the large-$k$ limits of the integral expressions given in \cite{berr2}. 
The Dirichlet and Neumann cases considered in \cite{berr3} are particular 
cases of Eq.~(\ref{eq:exam1}) corresponding to $a=0$ and $a=\pi /2$.

To show how to apply the semiclassical scheme for more general situations, 
we consider now a smooth potential barrier. 
In \cite{hell} an ensemble of random superpositions of Airy functions in $y$-direction 
and plane waves in $x$-direction is introduced,
which satisfies locally the Schr\"odinger equation for the potential $V(\vec{r})=Vy$. 
Ensemble average then gives, up to an overall constant, 
$\langle \psi^{r}(\vec{r})\partial_{y}\psi^{r}(\vec{r})\rangle=
\int_{0}^{\infty}{\rm Ai}\left[\Psi(y,Q)\right]{\rm Ai}'\left[\Psi(y,Q)\right]dQ$ 
with $\Psi(y,Q)=\left(V\hbar^{-2}\right)^{\frac{1}{3}}(y-y_{0})+\left(V\hbar^{-2}\right)^{
-\frac{2}{3}}Q^{2}$, where $y_{0}=e/V$ is the classical turning point. ${\rm Ai}(x)$ 
and ${\rm Ai}'(x)$ is the Airy function and its derivative  \cite{berr2,hell}. 
The classical paths (with no or one turning point)
required to construct the corresponding average via the semiclassical 
correlation function can be calculated in closed form as they are just parabolic flights. Using Eq.~(\ref{eq:Rsc}) and keeping only terms to leading order in $\hbar^{-1}$, 
we finally get  
\begin{eqnarray}
\label{eq:exam2}
\langle \psi(\vec{r})\partial_{y}\psi(\vec{r})\rangle=
\frac{2 m\delta e}{(2 \pi \hbar)^{3/2}N_{W}}\Gamma
\left(\frac{(2mV(y_{0}-y))^{1/2}\delta e}{\hbar V}\right) && \nonumber \\ 
\times \frac{1}{2\hbar \sqrt{2V}(y_{0}-y)^{2}}\sin 
\left(\frac{4 \sqrt{2mV}}{3\hbar}(y_{0}-y)^{3/2}\right), && \nonumber
\end{eqnarray}
which is again the asymptotic limit of the integral expression presented above.
Correspondingly, we recover the asymptotic limits of the results for the geometries
studied in \cite{hell2} in terms of a small number of paths.

To summarize, we showed how to efficiently treat wavefunction statistics for
closed systems by merging statistical with semiclassical concepts.
We demonstrated that all known (to us) results\cite{berr2,berr3,hell,hell2}
for specific, non-isotropc Random Wave Models 
are particular cases of the general approach presented here. 
It provides closed analytical 
expressions for statistical measures in terms of geometrical quantities
and builds the framework for incorporating arbitrary boundary conditions and 
confinement geometries.  

We thank S.\ Gnutzmann, G.~Foltin, P.~Schlagheck, M.\ Sieber, U.\ Smilansky,
and M.\ Turek for helpful conversations. This work was supported by the {\em 
Graduiertenkolleg} "Nonlinearity and Nonequilibrium in Condensed Matter" of 
the {\em Deut\-sche For\-schungs\-gemeinschaft}.

%
%


\begin{thebibliography}{10}

\vspace*{-1.5cm}

\bibitem{Im}
Y. Imry, {\it Introduction to Mesoscopic Physics} (Oxford University Press, New York 1997)

\bibitem{gut}
M. Gutzwiller {\it Chaos in Classical and Quantum Mechanics} (Springer, New York, 1990)

\bibitem{cambridge} See, e.g., the recent reviews by Kaplan and Heller, Fishman, and Mirlin
in {\em Supersymmetry and Trace Formulae}, ed.\ by I.V.\ Lerner, J.P.\ Keating,
and S.E.\ Khmelnitskii, (Kluwer, New York, 1999), and further references therein.

\bibitem{berr1}
M. V. Berry, {\it J. Phys. A} {\bf 10}, 2083 (1977)

\bibitem{mac}
S. W. McDonald and A. N. Kaufman, {\it Phys. Rev. Lett.} {\bf 42}, 1189 (1979)

\bibitem{coulomb-blockade-exp} A.~M.\ Chang et al., {\em Phys.~Rev.~Lett.}
{\bf 76}, 1695 (1996), J.~A.~Folk et al., {\em ibid} 1699 (1996)

\bibitem{note-CB}
Wave function correlations entering into 
interaction matrix elements influence spectral 
properties of quantum dots also beyond the independent-particle model
\cite{alhassid,mirlin,ahn,denn}.

\bibitem{alhassid} Y.~Alhassid {\em Rev.~Mod.~Phys. \bf 72}, 895 (2000)

\bibitem{mirlin}
A.~D. Mirlin, {\it Phys. Rep.~}{\bf 326}, 259 (2000)

\bibitem{ahn} K.-H.\ Ahn, K.\ Richter, and I.~H.\ Lee, 
{\em Phys.~Rev.~Lett.} {\bf  83}, 4144 (1999)

\bibitem{denn}
D. Ullmo, and H.~U. Baranger, {\it Phys. Rev. B~}{\bf 64}, 245324 (2001)  

\bibitem{sto}
See, e.g., Y. H. Kim, M. Barth, U. Kuhl, 
and H. J. St\"ockmann, cond-mat/0301411 and references therein.

\bibitem{uzy1}
G. Blum, S. Gnutzmann, and U. Smilansky, {\it Phys. Rev. Lett.} {\bf 88}, 114101 (2002)

\bibitem{har2}
E. E. Narimanov, H. U. Baranger, N.R. Cerruti, and S. Tomsovic, 
{\it Phys. Rev. B} {\bf 64},235329 (2001)

\bibitem{berr2}
M. V. Berry and H. Ishio, {\it J. Phys. A} {\bf 35}, L447 (2002)

\bibitem{berr3}
M. V. Berry, {\it J. Phys. A} {\bf 35}, 3025 (2002)

\bibitem{hell}
W. E. Bies and E. J. Heller, {\it J. Phys. A} {\bf 35}, 5673 (2002)

\bibitem{hell2}
W. E. Bies, N. Lepore, and E. J. Heller, {\it J. Phys. A} {\bf 36}, 1605 (2003)

\bibitem{prig}
V. N. Prigodin, {\it Phys. Rev. Lett.} {\bf 75}, 2392 (1995)

\bibitem{sied1}
M. Srednicki, {\it Phys. Rev. E} {\bf 54}, 954 (1996) 

\bibitem{sied2}
S. Hortikar, and M. Srednicki,  {\it Phys. Rev. Lett.} {\bf 80}, 1646 (1998)

\bibitem{note1}
The generalization to arbitrary dimensions and systems with broken time reversal symmetry 
is straight forward.

\bibitem{bee}
C.~W.~J.~Beenakker, Rev.~Mod.~Phys. {\bf 69}, 731 (1997) 

\bibitem{exp2}
See, e.g.,  A. M. Chang, H. U. Baranger, L.N. Pfeifer, and K.W. West, {\it Phys. Rev. Lett.} 
{\bf 73}, 2111 (1994)

\bibitem{bal}
R. Balian and C. Bloch, {\it Ann. Phys.} {\bf 60}, 401 (1970) 

\bibitem{pot}
A precise criterium is  
$|\vec{r}_{i}-\vec{r}_{j}|\ll a(\vec{r}_{i}+\vec{r}_{j})/2$
with $[a(\vec{r})]^{-1}=\nabla {\log}|V(\vec{r})|$.

\bibitem{cay}
F. Toscano and C. Lewenkopf, {\it Phys. Rev. E} {\bf 65}, 036201 (2002) 

\bibitem{yo}
J.~D.~Urbina and K.~Richter, nlin.CD/0304042 (2003)

\bibitem{note-finite-size}
This allows to roughly account for finite-size effects
using only direct paths as suggested in a numerical 
context \cite{denn}. 

\bibitem{Africa}
M. Robnik,{\it J. Phys. A} {\bf 17}, 1049 (1984) 

\bibitem{explain}
$R$ must satisfy $N_{W}\int 
R(\vec{r}_{i},\vec{r})R(\vec{r},\vec{r}_{j})d \vec{r}=R(\vec{r}_{i},\vec{r}_{j})$
for  consistency.
For this composition rule to hold, and to avoid serious normalization problems \cite{yo}
the window function $\Gamma(x)=\frac{\sin(x)}{x}$ is essential.

\bibitem{note-qm}
Quantum results 
for an energy interval $[E_{330},E_{370}]$
and mean level spacing 4 (for billiard area $A=1$).

\bibitem{uzy2}
M. Sieber {\em et.\ al,},
{\it J. Phys. A} {\bf 28}, 5041 (1995)  

\end{thebibliography}
\end{document}